\documentclass[12pt,draftclsnofoot,onecolumn]{IEEEtran}
\UseRawInputEncoding

\usepackage{lipsum}
\usepackage{optidef}
\usepackage{amsmath}
\usepackage{amsthm}
\usepackage{amssymb}
\usepackage{bm}
\usepackage{amsfonts}
\usepackage{paralist}
\usepackage{cite}

\theoremstyle{definition}
\newtheorem{rmk}{Remark}
\theoremstyle{definition}

\usepackage{caption}
\usepackage{graphicx}
\usepackage{xcolor}
\usepackage{subcaption}
\usepackage{algpseudocode}
\usepackage{algorithmicx}
\usepackage[ruled]{algorithm}
\usepackage{algpascal}
\usepackage{algc}
\usepackage[pagebackref=false,colorlinks,linkcolor=blue,citecolor=blue]{hyperref}

\makeatletter
\newcommand*{\rom}[1]{\expandafter\@slowromancap\romannumeral #1@}

\newcommand{\indep}{\perp \!\!\! \perp}
\makeatother
\begin{document}
	
	\title{ Direction-Based Jamming Detection and Suppression in mmWave Massive MIMO Systems}
	\author{ \ \\ Saeed Bagherinejad  and S. Mohammad Razavizadeh\\ \ \\
		Iran University of Science and Technology (IUST) \\ Narmak, Tehran 16846-13114, IRAN \\
	}
	
\maketitle
\begin{abstract}
In this paper, we study the problem of physical layer security in the uplink of millimeter-wave massive multiple-input multiple-output (MIMO) networks and propose a jamming detection and suppression method. The proposed method is based on directional information of the received signals at the base station antenna array. The proposed jamming detection method can accurately detect both the existence and direction of the jammer using the received pilot signals in the training phase. The obtained information is then exploited to develop a channel estimator that excludes the jammer’s angular subspace from received training signals. The estimated channel information is then used for designing a combiner at the base station that is able to effectively cancel out the deliberate interference of the jammer. By numerical simulations, we evaluate the performance of the proposed jamming detection method in terms of correct detection probability and false alarm probability and show its effectiveness when the jammer’s power is substantially lower than the user’s power. Also, our results show that the proposed jamming suppression method can achieve a very close spectral efficiency as the case of no jamming in the network.
\end{abstract}
\maketitle

\section{Introduction}
Recently, the advent of new mobile and wireless applications has caused a fast-growing demand for very high data rate communications in the fifth-generation (5G) and beyond wireless networks. There are many techniques that have been proposed to address these demands among them massive multiple-input multiple-output (MIMO) and millimeter-wave (mmWave) are the most promising \cite{marzetta_larsson_yang_ngo_2016, Survay_mmWave}. In addition to higher bandwidth in the mmWave bands, their shorter wavelength enables us to employ a large number of antennas at the base stations (BSs) and makes it appropriate for combination with the massive MIMO systems \cite{Emil2017Book}.
One of the inherent weaknesses of wireless networks is their vulnerability to security attacks at the physical channels including jamming and eavesdropping. Security can be provided in different layers of the network in which the physical layer security is a powerful technique that has attracted much attention in the recent years\cite{PLS_Survay_2018}. Massive MIMO systems are naturally immune to a \emph{passive eavesdropping} attacks due to their ability to create very narrow beams toward the legitimate users which reduces any signal leakage to the illegitimate terminals. However, an \emph{active eavesdropping} attack that disrupts the training phase by transmitting a jamming signal can reduce the secrecy rate of the massive MIMO systems \cite{PLS_Kapetanovic_2015,Basciftci2015_SecMaMIMO}. In addition to the training phase, a jamming attack can also occur at the data transmission phase with the goal of decreasing spectral efficiency of the system \cite{Pirzadeh_Subv_2016}. Consequently, one of the crucial problems in the massive MIMO systems is detecting jamming attacks and then using techniques to suppress them or alleviate their effects. This is of high importance especially for operation in hostile environments. 

The problems of jamming detection and suppression are discussed in many of the prior works in the field of physical layer security in massive MIMO systems \cite{Kapetanovic_Random_Training_2013,Kapetanovic_Det_2014,PilotSpoofing_2019,Vinogradova_Matrix_theory_2016,Nie_Matrix_Theory_2017,akhlaghpasand2017WCL,Wang_PSA_2019,Xu_2020_PSA,Zhao_IA_jamm_supp_2017,akhlaghpasand2019TCAS,Do2018_JamRes,Akhlagh_2020}. In \cite{Kapetanovic_Random_Training_2013}, a jamming attack detection scheme is proposed based on pilots drawn randomly from a known constellation. The authors in \cite{Kapetanovic_Det_2014,PilotSpoofing_2019} use additional signaling and cooperation between the BS and users to suggest jamming attack detection methods. Two approaches based on random matrix theory for jamming detection have been proposed in \cite{Vinogradova_Matrix_theory_2016,Nie_Matrix_Theory_2017} in which the authors use the rank of the received signal's covariance matrix and eigenvalues of the received signal matrix. Also in \cite{akhlaghpasand2017WCL}, the intentionally unallocated pilots are leveraged by the authors to propose a jamming detection technique. Authors of \cite{Wang_PSA_2019,Xu_2020_PSA} propose two similar approaches based on double channel training where the received signals during two training phases in a single coherence block are compared to determine the presence of pilot spoofing attack. None of the aforementioned researches or other related works utilize directional information for detecting the presence of the jamming attack. On the jamming suppression problem, a jamming attack mitigation method is proposed in \cite{Vinogradova_Matrix_theory_2016} in which, the users' eigen-subspace is estimated and then the received signal are projected onto this subspace. Zhao et. al. in \cite{Zhao_IA_jamm_supp_2017} propose a jamming rejection technique using cooperation between transmitter and receiver and using an interference alignment technique. In \cite{akhlaghpasand2019TCAS} and \cite{Do2018_JamRes}, two approaches are suggested that leverage unused pilots to estimate the jammer's channel information and then utilize that information to null out the jammer's signal. Also, a framework for estimating legitimate users' channels and maximizing spectral efficiency under jamming attack is proposed in \cite{Akhlagh_2020} which is based on statistical information of the channels and can be implemented in spatially correlated channels.

High angular resolution is one of the most important properties of the massive MIMO systems which can be used for combating different challenges of these networks \cite{PilotDecont_Gong_2019,Xie2017TVT}. Accurate directional information provided by high angular resolution can be also utilized to mitigate the undesirable effects of the jamming and eavesdropping attacks \cite{SecTrmmWaveYingJu2017,Secure_Beam_Domain_2018,Airborne_2019,Xu_mmWave_PLS_2020}. For example, a secure downlink transmission scheme exploiting angular information is suggested in \cite{SecTrmmWaveYingJu2017} where authors assume the BS has perfect knowledge of the eavesdropper's directional information and propose three precoding methods based on this information. In \cite{Secure_Beam_Domain_2018}, a beam domain secure transmission is proposed which optimizes allocated power to each of angular paths to maximize the secrecy sum-rate of the downlink transmission in a massive MIMO system where a passive multi-antenna eavesdropper is present. Authors in \cite{Airborne_2019} propose a technique to estimate the angular information of users and the eavesdropper in airborne massive MIMO systems and use this information to enhance channel estimation performance.
Xu et. al. in \cite{Xu_mmWave_PLS_2020} suggest a hybrid beamforming technique that transmits confidential data towards a legitimate user's dominant directions and an artificial noise signal towards all other directions using the statistical angular information. All the abovementioned works that rely on directional information have assumed that the BS is aware of the presence of the adversarial terminal, therefore, they do not discuss the detection of the adversary. Also, in none of the above and other related works, the problem of suppressing a jamming attack on the uplink transmission of a massive MIMO system using directional information is addressed.
 
Motivated by the above, in this paper, we address the problem of jamming detection and suppression in mmWave massive MIMO networks using \emph{directional (angular) information} of the users and jammer. The network consists of a BS with a large number of antennas that serves a number of single-antenna users. There is also a jammer in the network that transmits interfering signals in both training and data transmission phases to sabotage the system’s performance. A discrete channel model comprising a number of spatially \emph{resolvable paths} (RP) is presented to describe the mmWave massive MIMO channel in the angular domain. Because of the limited scattering in the environment, the received signal from each terminal (i.e. a user or the jammer) arrives the BS array from only a few RPs which are referred to as \emph{active} RPs of that terminal.
First, for the jamming detection purpose, based on the received pilot signals in the training phase, we propose a method that estimates the set of RPs through which a particular pilot is received. Then, we check whether there are active RPs that are common among a large number of pilots and accordingly detect the presence as well as the directional information of the jammer. This information along with other information obtained about the users' directional information is then used for jamming suppression in the next phase. In the next part of the paper and for the jamming suppression purpose, utilizing the above directional information, we propose a channel estimation scheme which is based on projecting received pilot signals onto the orthogonal complement of the jammer’s angular subspace. Then, the estimated channel information is used for designing the combining vectors which are orthogonal to the jammer’s channel and cancels out the jamming signal. Note that the directional information obtained in the channel training phase can be used for the jamming suppression in several consecutive intervals of coherence time. The reason is that the spatial characteristics, such as active RPs of a terminal's channel, change slower than the small scale fading parameters such as the gains along with the active RPs. The key contributions of this paper can be summarized as follows:

\begin{itemize}
	\item A jamming detection and suppression method in a mmWave massive MIMO network is proposed which relies on directional (angular) information obtained during the channel training phase. Received energy along with each RP over several sub-carriers (sub-channels) is leveraged to determine if a RP is an active RP or not. 
	\item In contrast to other works, our proposed jamming detection technique does not require any prior knowledge of the users’ and jammer’s channel state information or any necessary precondition such as unused pilots in the network. Also, it does not introduce any additional signaling overhead.
	\item The proposed detection technique is capable of detecting the jammer even if its power is very low. Besides, the false alarm probability of this detection method is substantially lower than other similar techniques. The performance of the proposed scheme can be further enhanced  by increasing the number of antennas at the BS or the number of sub-carriers (sub-channels) at the system. The jamming detection method directly yields directional information of the jammer which can be utilized to suppress the jamming attack over several intervals of coherence time.
	\item The proposed jamming suppression also can efficiently cancel out the jamming attacks. We show that this method performs very well and similar to the case where no jamming  is present in the network. Also, its performance does not depend on the jammer’s power and is effective even with relatively high-power jamming attacks.
\end{itemize}

\begin{figure}[t]
		\centering
		\includegraphics[width=8.5cm]{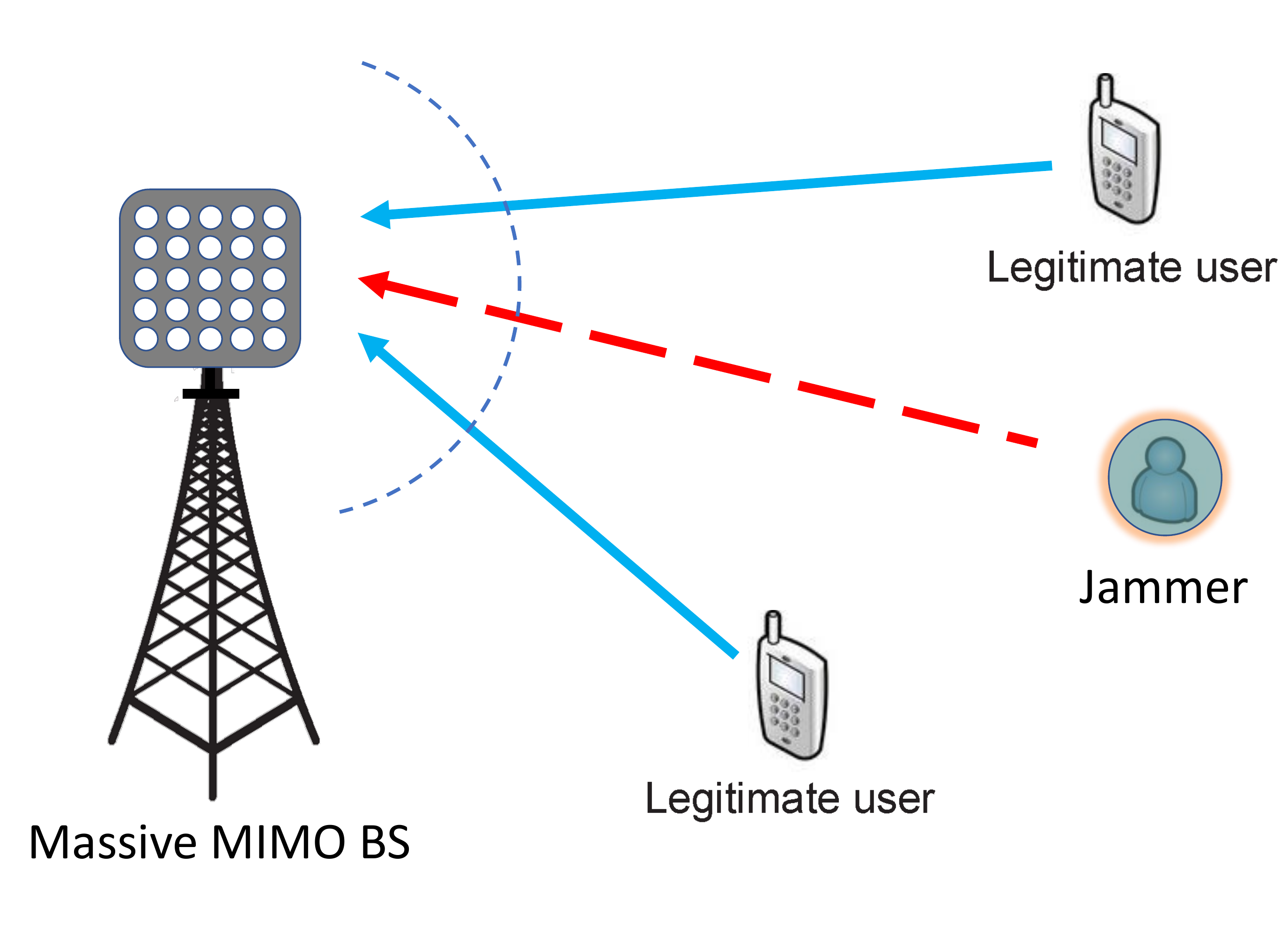}
		\caption{System model of a massive MIMO network attacked by a jammer.} \label{FigSysMod1}
\end{figure}

\section{System Model}\label{Sys_Mod}
	As depicted in Fig. \ref{FigSysMod1}, we consider uplink of a single-cell  massive MIMO system consisting of a BS with a large array of $M$ antennas that serves $K$ single-antenna legitimate users in the presence of a single-antenna jammer, which all are randomly located in the cell. The network operates in the mmWave bands and a multi-carrier transmission with $N$ sub-carriers (or sub-channels) is considered. We also consider a block-fading channel model where channel gains are constant within a coherence block of length $T=T_cN_c$ symbols and varies  independently from one block to another. $T_c$ and $N_c$ denote the number of time symbols in a coherence time and the number of sub-carriers in a coherence bandwidth, respectively. 

\subsection{Angular Domain Channel Modeling}\label{channel_model}
	Since our proposed scheme is based on the directional information of the channels, we need a channel model that properly describes signals in the angular domain. We assume an uniform linear array (ULA) at the BS and for each direction of arrival (DOA), $\theta\in [-\pi/2,\pi/2]$, the BS has the following array steering vector
	\begin{align}\label{AMV}
	\nonumber \mathbf{a}(\theta)=\frac{1}{\sqrt{M}}[1,e^{-j\frac{2{\pi}d}{\lambda}sin(\theta)},e^{-j2\frac{2{\pi}d}{\lambda}sin(\theta)},
	\\\ldots
	,e^{-j(M-1)\frac{2{\pi}d}{\lambda}sin(\theta)}]^T,
	\end{align}
	where $d$ and $\lambda$ are array element spacing and carrier wavelength, respectively. It is assumed that $d=\frac{\lambda}{2}$. Also, $\Theta=sin(\theta)$ denotes the \textit{directional sine} and since $\theta$ is in the interval of $[-\pi/2,\pi/2]$, there is a one-to-one mapping between $\theta$ and $\Theta$. We assume that the DOAs from which signals of the $k$th user are received  are uniformly distributed in the interval of  $\mathcal{I}_k=[\bar{\theta}_k-\Delta_k/2,\bar{\theta}_k+\Delta_k/2]$, where $\bar{\theta}_k$ and $\Delta_k$ are average incident angle and angular spread, respectively. We call the interval $\mathcal{I}_k$ as the \emph{angular span} of the $k$th user.
	
	The angular resolution of massive MIMO depends on the \emph{array's length},  $L$ which is equal to $L=M\frac{d}{\lambda}$ \cite{Tse2005}. If the directional sine of two paths differ less than $1/L$, these paths would not be resolvable by the array \cite{Tse2005}. Hence, the angular domain can be sampled at fixed angles with a spacing of $1/L$ in the directional sine. The steering vectors with angular spacing of $1/L$ at their $\Theta$ are orthonormal, thus they can form an orthogonal basis for channel expansion that can be represented as \cite{ExpStuMIMO2007}
	\begin{equation}\label{Orth_Basis}
	\mathbf{U}\triangleq [\mathbf{a}(\phi_1),\mathbf{a}(\phi_2)),\ldots,\mathbf{a}(\phi_{M})],
	\end{equation}
	where $\phi_i\triangleq sin^{-1}( \frac{1}{L}(i-1-(M-1)/2)),  i=1,2,\ldots,M$ are the sampled angles. 
	
	By these definitions, we can decompose the channels to $M$ different physical directions which we refer to them as resolvable path (RP). The gain of the $i$th RP can be represented as the superposition of all paths whose $\Theta$ are located within a window of width $1/L$ around $sin(\phi_i)$. Subsequently, we define a virtual channel representation (VCR) model which samples the angular domain using spatial orthogonal basis $\mathbf{U}$. The VCR channel vector of the $k$th user in the $n$th sub-carrier can be denoted by
	\begin{equation}\label{Channel_Model}
	\mathbf{h}_k^n=\sqrt{\frac{M\beta_k}{C_k}}\mathbf{U}\tilde{\mathbf{g}}_k^n,\; n=1,\ldots,N ;\; k=1,\ldots,K,
	\end{equation}
	where  $\tilde{\mathbf{g}}_{k}^{n}=[\tilde{g}_{k,1}^{n},\tilde{g}_{k,2}^{n},\ldots,\tilde{g}_{k,M}^{n}]$ is a complex gain vector and $\tilde{g}_{k,i}^{n}$ indicates the small scale gain of $i$th RP. If a RP is located in the angular span of the $k$th user, i.e. $\phi_i\in \mathcal{I}_k$, we call it as an active RP of that user. Since each active RP includes several physical paths, we assume that the small scale random modelling of an active RP is a complex Gaussian random coefficient with zero mean and unit variance, i.e. $\tilde{g}_{k,i}^{n} \sim \mathcal{CN}(0,1)$. If a RP is not an active RP, i.e. $\phi_i\notin\mathcal{I}_k$, its gain would be equal to zero, i.e. $\tilde{g}_{k,i}^{n} =0$. Furthermore, $\beta_k$ and $C_k$ are large scale fading coefficient and the number of active RPs of the $k$th user, respectively. Due to limited scattering characteristic of the mmWave channels, the received power is concentrated on a narrow interval in the angular domain. Thus, the angular spread is relatively small and only a few RPs are located within it. Thus, $C_k$ is considerably smaller than the number of array antennas, i.e. $C_k\ll M$ \cite{SecTrmmWaveYingJu2017}.
	
	The angular span of each user is constant over different sub-carriers since the spatial propagation characteristics are unchanged within system's bandwidth. Therefore
	\begin{equation}\label{Common_Sparsity}
	\Omega_k=supp({\tilde{\mathbf{g}}_k^1})=supp(\tilde{\mathbf{g}}_k^2)=\ldots=supp(\tilde{\mathbf{g}}_k^N),
	\end{equation}
	where $supp{}$ denotes the support set of a vector. This equation implies the active RPs of each user are unchanged over different sub-carriers. The set of $\Omega_k$ is also referred to as \emph{spatial signature}.

\begin{figure}[t]
	\centering
	\includegraphics[width=8cm]{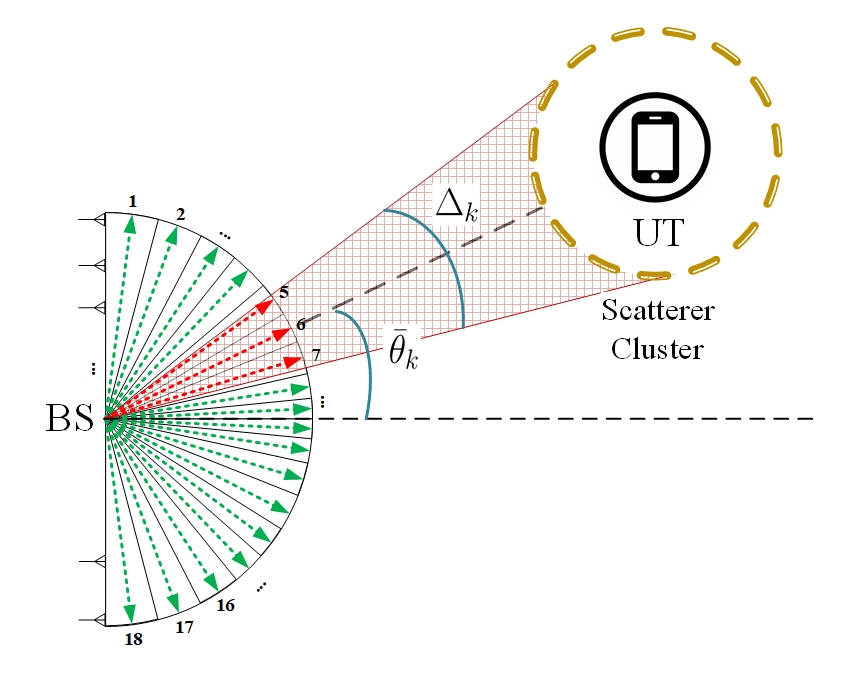}
	\caption{An example of a virtual channel representation (VCR)  model for an array with $M=18$ antennas.} \label{FigChannelMod}
\end{figure}

	Fig. \ref{FigChannelMod} illustrates an example of the described channel model in angular domain for $M=18$ antennas at the BS. Solid arrows show the spatially resolvable paths that three of them are located within the angular span of the user and in fact are the active RPs. These active RPs are demonstrated by red arrows. Thus, in this example, the set of active RPs for this user is $\Omega_k = \{5,6,7\}$.
	
	Similar to the users' channels, the jammer's channel in the $n$th sub-carrier can also be denoted by $\mathbf{h}_w^n=\sqrt{{M\beta_w}/{C_w}}\mathbf{U}\tilde{\mathbf{g}}_w^n$ where $\beta_w$ and $C_w$ denote the large scale fading coefficient and the number of active RPs of the jammer, respectively. The angular span of jammer is also denoted by $\mathcal{I}_w=[\bar{\theta}_w-\Delta_w/2,\bar{\theta}_w+\Delta_w/2]$, where $\bar{\theta}_w$ and $\Delta_w$  are the  average incident angle and  angular spread of the jammer, respectively. Also, similar to \eqref{Common_Sparsity}, the set of indices of active RPs (spatial signature) for the jammer is denoted by $\Omega_w$.
	
	We assume that the users and the jammer are randomly distributed in the cell, thus we consider average incident angle of the $k$th user and jammer to be uniformly distributed in the intervals of  $[-\pi/2+\Delta_k/2,\pi/2-\Delta_k/2]$ and $[-\pi/2+\Delta_w/2,\pi/2-\Delta_w/2]$, respectively. The $k$th user's and the jammer's channels can be alternatively defined as
	\begin{align}
	\label{Channel_Model_User_sum}\mathbf{h}_k^n=\sqrt{\frac{M\beta_k}{C_k}}\sum_{i\in\Omega_k}\tilde{g}_{k,i}^{n}\mathbf{a}(\phi_i)\\ \mathbf{h}_w^n=\sqrt{\frac{M\beta_w}{C_w}}\sum_{i\in\Omega_w}\tilde{g}_{w,i}^{n}\mathbf{a}(\phi_i) 
	\end{align} 
	
	\subsection{Pilot and Data Transmission Phases}
	Transmission from the users to the BS in the uplink occurs in two phases. First, in the channel training pahse, each user sends its allocated orthogonal pilot sequence to the BS for estimating that user's channel. In the second phase which is data transmission phase, the user sends its data to the BS. The jammer also transmits jamming signals during both training and data phases with different powers $q_t$ and $q_d$, respectively.
	
	\subsubsection{Channel Training}
	Assuming a block fading channel with $N_c$ sub-channels that are placed within its coherence bandwidth. As a result, only one sub-carrier out of $N_c$ needs to be estimated using orthogonal pilot sequences \cite{Emil2017Book,marzetta_larsson_yang_ngo_2016}. We denote the number of estimated sub-carriers by $N_e=N/N_c$. The set of orthogonal pilots is $\mathbf{S}=[\mathbf{s}_1,\mathbf{s}_2,\ldots,\mathbf{s}_\tau]\in \mathbb{C}^{\tau\times\tau}$, where $\tau$ is length of orthogonal pilots, and we assume $\tau=K$. Also, pilots are considered to have unit power, i.e. $\mathbf{S}^H\mathbf{S}=\mathbf{I}$. The $k$th user transmits the $k$th pilot over $N_e$ predefined sub-carriers. At this phase, the received signal in the $n$th sub-carrier can be represented as
	\begin{equation}\label{Rec_Sig_Pilot}
	\mathbf{Y}_t^n=\sum_{k=1}^{K}\sqrt{\tau p_{t,k}}\mathbf{h}_k^n\mathbf{s}_k^H+\sqrt{\tau q_t}\mathbf{h}_w^n(\mathbf{s}_w^n)^H+\mathbf{Z}_t^n,
	\end{equation}
	where $p_{t,k}$ is the pilot transmission power of the $k$th user and $\mathbf{s}_w^n\in\mathbb{C}^{\tau\times1}$ is the jamming pilot signal sent by the jammer in the $n$th sub-carrier. $\mathbf{Z}_t^n\in\mathbb{C}^{M\times\tau}$ is also additive white Gaussian noise (AWGN) whose elements are $[\mathbf{Z}_t^n]_{i,j}\sim\mathcal{CN}(0,\sigma_z^2)$. In order to estimate the $k$th user's channel, the BS de-spread the received signal in \eqref{Rec_Sig_Pilot} with the pilot $\mathbf{s}_k$ as
	\begin{equation}\label{Rec_Time_Pilot}
	    \mathbf{y}_{t,k}^n=\mathbf{Y}_t^n\mathbf{s}_k=\sqrt{\tau p_{t,k}}\mathbf{h}_k^n+\sqrt{\tau q_t}\gamma_k^n\mathbf{h}_w^n+\mathbf{z}_{t,k}^n,
	\end{equation}
	where $\mathbf{z}_{t,k}^n=\mathbf{Z}_t^n\mathbf{s}_k$ is the correlation between noise and  the $k$th pilot. Also, $\gamma_k^n=(\mathbf{s}_w^n)^H\mathbf{s}_k$  indicates the correlation between the jamming signal and the $k$th pilot. It is clear that the interference exerted by other users have been eliminated by orthogonality of pilots; however, the deliberate interference by the jammer signal still remains.
	
	A jammer usually aims to limit the SE of the massive MIMO network rather than targeting a special user. The reason is that, in the massive MIMO systems, if a random pilot hopping method is used for assigning pilots to the users \cite{akhlaghpasand2019TCAS,Do2018_JamRes,Basciftci2015_SecMaMIMO}, it is difficult for a jammer to acquire a particular user's pilot. However, it is reasonable to assume that the jammer has knowledge about the pilot set $\mathbf{S}$.
	
	In addition, a necessary condition for jamming signal with the purpose of mitigating performance of the system, is that the $Prob(\gamma_k^n=0)=0$ which can be achieved by spreading jamming power over all the pilots. A known method for designing jamming signal in the training phase is that $E(|(\mathbf{s}_w^n)^H\mathbf{s}_i|^2)=1/\tau$. One of the signals that satisfies this property is \cite{akhlaghpasand2019TCAS}
	\begin{equation}\label{Jamming_Signal}
	\mathbf{s}_w^n\thicksim\mathcal{CN}(0,\frac{1}{\tau}\mathbf{I}),\;n=1,\ldots,N_e.
	\end{equation}
	
	To restrain the jamming attack on the training phase, some extra information about the users' and jammer's channels is needed which is the directional information in our work. In Sec. \ref{DOA_EST} and \ref{JAMMING_DET}, we will investigate how to obtain this information and then utilize it in Sec. \ref{JAMMING_SUP} to propose the channel estimation technique.
	
	\subsubsection{Data Transmission}
	In the uplink data transmission phase, each user $k$ sends its data with power $p_{k,d}$. Simultaneously, the jammer sends its jamming signal to the BS. The received signal at the BS over the $n$th sub-carrier can be expressed as, 
	\begin{equation}\label{Data_Sig_Rec}
	\mathbf{y}_d^n=\sum_{k=1}^{K}\sqrt{p_{d,k}}\mathbf{h}_k^nx_k^n+\sqrt{q_d}\mathbf{h}_w^nx_w^n+\mathbf{z}_d^n
	\end{equation}
	where $\mathbf{z}_d^n \thicksim \mathcal(\mathbf{0},\sigma_z^2\mathbf{I})$ is the AWGN noise, $x_k^n$ and $x_w^n$ are the $k$th user's data and jamming signal, respectively. These signals are considered to be i.i.d Gaussian random variable with zero mean and unit variance, i.e. $x_k^n,x_w^n\sim\mathcal{CN}(0,1)$. Then the BS receives this signal and decodes it by the use of information obtained in the training phase. Suppose $\mathbf{V}^n$ is the decoding (combining) matrix at the BS and $\mathbf{v}_k^n$ is its $k$th column which is the receive combining vector of the $k$th user. By calculating the inner product of $\mathbf{v}_k^n$ and $\mathbf{y}_d^n$, the signal from the $k^{th}$ user in the $n$th sub-carrier can be decoded as
	\begin{align}\label{Data_Sig_Time_Comb}
	\nonumber y_{d,k}^n={\mathbf{v}_k^n}^H\mathbf{y}_d^n&=\sqrt{p_{d,k}}{\mathbf{v}_k^n}^H\mathbf{h}_k^nx_k^n+\sum_{\stackrel{l=1}{l\neq k}}^{K}\sqrt{p_{d,l}}{\mathbf{v}_k^n}^H\mathbf{h}_l^nx_l^n\\
	&+\sqrt{q_d}{\mathbf{v}_k^n}^H\mathbf{h}_w^nx_w^n+{\mathbf{v}_k^n}^H\mathbf{z}_d^n.
	\end{align}
	Since the BS has an imperfect CSI of user's channel, we utilize a lower bound of the capacity called \textit{use-and-then-forget} which is widely used in massive MIMO \cite{marzetta_larsson_yang_ngo_2016,Emil2017Book}. The name originates from the fact that estimated channels are used for the design of the receive combining vectors and then disregarded in the signal detection. This achievable rate is defined as follows
	\begin{align}\label{Ach_Rate}
	R_k^n=(1-\frac{\tau}{T})\mathbb{E}_{\Psi}\{\log_2(1+\rho_k^n)\},
	\end{align}
	where $\rho_k^n$ is the effective SINR  defined in (\ref{SINR}) at the top of the next page. Note that this SINR is conditioned on  $\Psi$ which is the directional information and will be discussed in the following sections.
	\begin{figure*}[t]
		\normalsize
		\begin{align}\label{SINR}
		\rho_k^n=\dfrac{p_{d,k}|\mathbb{E}\{{\mathbf{v}_k^n}^H\mathbf{h}_k^n|\Psi\}|^2}{p_{d,k}\mathtt{var}\{{\mathbf{v}_k^n}^H\mathbf{h}_k^n|\Psi\}+\sum_{\stackrel{l=1}{l\neq k}}^{K}p_{d,l}\mathbb{E}\{|{\mathbf{v}_k^n}^H\mathbf{h}_l^n|^2|\Psi\}+q_d\mathbb{E}\{|{\mathbf{v}_k^n}^H\mathbf{h}_w^n|^2|\Psi\}+\mathbb{E}\{|{\mathbf{v}_k^n}^H\mathbf{z}_d^n|^2|\Psi\}}
		\end{align}
		\hrulefill
		\vspace*{4pt}
	\end{figure*}
	
	\section{Jamming Detection}
    In this section, we discuss the proposed jamming detection scheme which consist of two stages. In the first stage, the signals in (\ref{Rec_Time_Pilot}) are used to estimate the active RPs from which any particular pilot is received. To this end, we exploit the received training signals in several sub-carriers to distinguish  the active RPs (i.e those that include a signal) from the RPs that only include noise. Then, in the second stage, we present our proposed detection method. In this method,  after determining the active RPs, we evaluate the number of pilots that are received through each RP and in this way we determine if there is a jammer in the network.
	\subsection{Active RP Estimation}\label{DOA_EST}
	As we explained before, the $k$th pilot is transmitted by both the $k$th user and the jammer. The union set of active RP indices for the jammer and the $k$th user is defined as $\Omega_{k,w}=\Omega_k\cup\Omega_w$. In fact, this set shows the spatial directions from which the $k$th pilot is received.
		
	Calculating the inner product of $\mathbf{U}$ in \eqref{Orth_Basis} with $\mathbf{y}_k^n$ in (\ref{Rec_Time_Pilot}) maps the received signals to the angular domain. Therefore,
	\begin{equation}\label{Rec_Ang_Dom}
	\tilde{\mathbf{y}}_{t,k}^n=\mathbf{U}^H\mathbf{y}_{t,k}^n=\sqrt{\tau p_{t,k}\mu_k}\tilde{\mathbf{g}}_k^n+\sqrt{\tau q_t\mu_w}\gamma_k^n\tilde{\mathbf{g}}_w^n+\tilde{\mathbf{z}}_{t,k}^n,
	\end{equation}
	where $\tilde{\mathbf{z}}_k^n=\mathbf{U}^H\mathbf{z}_k^n$ whose distribution is the same as $\mathbf{z}_k^n$ since $\mathbf{U}$ is an unitary matrix, i.e. $\tilde{\mathbf{z}}_k^n\thicksim \mathcal{CN}(\mathbf{0},\sigma_z^2\mathbf{I})$. We need a sensing scheme that can distinguish between the active RPs and the RPs that only consist of noise. For this purpose, we first define a hypothesis test for the $i$th RP as
	\begin{align}\label{RP_Det}
	    \nonumber&\mathcal{H}_{0,i}^{RP}: [\tilde{\mathbf{y}}_{t,k}^n]_i=[\tilde{\mathbf{z}}_{t,k}^n]_i\\
	    &\mathcal{H}_{1,i}^{RP}: [\tilde{\mathbf{y}}_{t,k}^n]_i= \begin{cases}
	    \sqrt{\tau p_{t,k}\mu_k}\tilde{g}_{k,i}^n+\sqrt{\tau q_t\mu_w}\gamma_k^n\tilde{g}_{w,i}^n+[\tilde{\mathbf{z}}_{t,k}^n]_i\\
	    \sqrt{\tau p_{t,k}\mu_k}\tilde{g}_{k,i}^n+[\tilde{\mathbf{z}}_{t,k}^n]_i\\
	    \sqrt{\tau q_t\mu_w}\gamma_k^n\tilde{g}_{w,i}^n+[\tilde{\mathbf{z}}_{t,k}^n]_i
	\end{cases}
	\end{align}
	where $\mathcal{H}_{1,i}^{RP}$ denotes three cases of (a) both the $k$th user and the jammer transmit the signal, (b) the user transmits the signal and (c) the jammer transmit the signal all along with the $i$th RP. $\mathcal{H}_{0,i}^{RP}$ denotes the opposite case in which no signal is received from the jammer or the $k$th user and only a noise signal is present along with the $i$th RP.
	
	Furthermore, the energy received along a certain RP can be utilized to determine if it is a active RP or not. This stems from the fact that receiving signal along a RP increases the energy level comparing to the case that it only receiving noise. Thus, it is judicious to infer that a RP is active if its energy is more than a predefined threshold. To this end, for the $i$th RP, we compute the sum received energy  from $N_d$ sub-carriers as
	\begin{equation}\label{Sum_Over_sub-carrier}
	    W_{i,k}\triangleq\sum_{n=1}^{N_d}|[\tilde{\mathbf{y}}_{t,k}^n]_i|^2.
	\end{equation}
	After computing the received energy, we can use it for determining the active RPs according to what is mentioned above. Therefore, we define the decision rule as 
	\begin{equation}\label{RP_Dec_Rule}
	    W_{i,k}\underset{\mathcal{H}_{0,i}^{RP}}{\overset{\mathcal{H}_{1,i}^{RP}}{\gtrless}}\epsilon_k,
	\end{equation}
	where $\epsilon_k$ is the decision threshold which can be chosen to satisfy various criteria. Here, we propose a condition for selecting the threshold that guarantees the probability of false alarm (i.e. declaring a non-active RP as an active RP) to be less than a given value $\eta$. The condition is defined as 
	\begin{equation}\label{Threshold_Cond}
		Prob\{W_{i,k}>\epsilon_k|\mathcal{H}_{0,i}^{RP}\}\leq\eta.
	\end{equation} 
    By choosing $\eta$ close to zero, e.g. $10^{-3}$, and determining $\epsilon_k$ accordingly, it will be assured that the RPs selected by the decision rule in (\ref{RP_Dec_Rule}) are always included in $\Omega_{k,w}$. Note that, $W_{i,k}$ when the $i$th RP only comprises of noise, is a Gamma distributed random variable, i.e $W_{i,k}\thicksim Gamma(N_d,\sigma_z^2)$, which can be used to calculate $\epsilon_k$.
	
	In order to efficiently estimate the active RPs, the user's and jammer's powers have to be sufficiently large compared to the noise power. It is rational to assume that the users' powers have been adjusted to be adequately more than noise power. However, there is no control over the jammer's power. In the case of low power jamming, the BS can use a larger number of sub-carriers. It enhances the RP detection performance since the gains of active RPs in different sub-channels are independent and it is highly probable that at least some of these gains are large enough to make the RP detectable. In general, increasing the number of received samples in the summation (16) would always help to improve the detection performance. It can be done either by increasing the number of considered sub-channels or considering the received training signal over several consecutive intervals of coherence time since the directional characteristics of channels change more slowly in time. This means we can sum the energy received along a RP over several sub-channels and in some consecutive intervals of time coherence. The impact of increasing number of sub-carriers or number of coherence time intervals is the same, thus, we only consider one coherence time and adjust the number of sub-channels.
	
	Finally, the estimated set of RPs for $k$th pilot is described as follows
	\begin{equation}\label{Est_RP_Set}
	    \hat{\Omega}_{k,w}=\{i\in\{1,\ldots,M\}| W_{i,k}>\epsilon_k\}.
	\end{equation}
	Having estimated the active RP sets, we exploit them to suggest a jamming detection scheme in the next section.
	
	\subsection{Jamming Detection}\label{JAMMING_DET}
	In this section, we propose a jamming detection scheme based on the estimated RP sets in the previous section. Initially, we discuss a principal characteristic of users' active RP sets $\{\Omega_1,\Omega_2,\ldots,\Omega_K\}$, which we are going to use to propose a jamming detection method.
	
	\begin{rmk}
		Suppose that average incident angles of all $K$ user are uniformly distributed in the interval of $(\frac{\pi}{2}-\frac{\Delta}{2},-\frac{\pi}{2}+\frac{\Delta}{2})$, and the angular spread is equal for all user, i.e. $\Delta_1=\ldots=\Delta_K=\Delta$. Then, the probability that $g$ out of $K$ users have at least one common active path,  $P_g^K$, is upper-bounded by
		\begin{equation}\label{Eq12}
		P_g^K\leqslant\bar{P}_g^K,
		\end{equation}
		where $\bar{P}_g^K=\min(1,\binom{K}{g}(1-(\frac{\pi-2\Delta}{\pi-\Delta})^2)^{g-1})$.
		\begin{proof}
			Please refer to \textbf{Appendix \ref{APP1}}.
		\end{proof} 
	\end{rmk}
    It should be noted that when $g$ has a low value, e.g. $g=2$ the upper-bound is close to one, i.e. $\bar{P}_g^K \sim 1$. However, as $g$ increases to $K$, $\bar{P}_g^K$ exponentially decreases to a very low value. For instance, for $K=10$, $\Delta=\pi/18$ and $g=6,8,10$, the upper-bound is proportional to $\bar{P}_6^{10}\sim 10^{-3},\;\bar{P}_8^{10}\sim10^{-5}$ and $\bar{P}_{10}^{10}\sim10^{-9}$. Thus, it is rational to assume that it is not likely for a large number of users to have common active RPs. Another conclusion that can be drawn is that a common active RP  can indicate the existence of a jammer.
	
	Having this property, we define another hypothesis test as 
	\begin{align}\label{Eq14}
	    \nonumber&\mathcal{H}_0^{JD}: \mathbf{Y}_t^n=\sum_{k=1}^{K}\sqrt{\tau p_{t,k}}\mathbf{h}_k^n\mathbf{s}_k^H+\mathbf{Z}_t^n \\
	    &\mathcal{H}_1^{JD}: \mathbf{Y}_t^n=\sum_{k=1}^{K}\sqrt{\tau p_{t,k}}\mathbf{h}_k^n\mathbf{s}_k^H+\sqrt{\tau q_t}\mathbf{h}_w^n(\mathbf{s}_w^n)^H+\mathbf{Z}_t^n,
	\end{align}
	where the hypothesis $\mathcal{H}_0^{JD}$ and $\mathcal{H}_1^{JD}$ denote absence and presence of the jammer, respectively. Then, we define the following set that contains indices of common active RPs between at least $g$ pilots
	\begin{equation}\label{Eq13}
	    \mathcal{Q}_g=\{i\in\{1,\ldots,M\}|\mathcal{R}(i)\geq g\},\; g=2,\ldots,K
	\end{equation}
	where $\mathcal{R}(i)$ is a function that returns the number of pilots that is received along $i$th RP according to the estimated sets $\{\hat{\Omega}_{1,w},\hat{\Omega}_{2,w},\ldots,\hat{\Omega}_{K,w}\}$. In addition, as illustrated in \textbf{Remark 1}, in the absence of the jammer, we anticipate only a small subset of pilots to have common active RPs since then pilots would be only received from the users' RPs. On the other hand, when a jammer is present in the network, its active RPs would be common for many of the pilots and hence, the number of pilots with common RPs would be relatively large. Based on this fact, we infer that if the number of pilots that have common RPs is more than a predefined threshold, a jammer is present in the network and vice versa. Therefore, we can rewrite the detector as
	\begin{eqnarray}\label{Eq15}
	\nonumber\mathcal{H}_0^{JD}: \mathcal{Q}_g=\emptyset \\
	\mathcal{H}_1^{JD}: \mathcal{Q}_g\neq\emptyset.
	\end{eqnarray}
	This decision rule states that if $\mathcal{Q}_g$ is an empty set where $g$ is the predefined threshold, the number of pilots which have common RPs is less than $g$ and thus, there is no jammer present in the network. On the other hand, when $\mathcal{Q}_g$ is a non-empty set, there are active RPs common between at least $g$ pilots and therefore, there is a jammer present in the network.
	Furthermore, the parameter $g$ should be selected properly to avoid a high false alarm probability (FAP). If $g$ is not large enough, the set $\mathcal{Q}_g$ can be non-empty even when there is no jammer, which leads to a high FAP. If $g$ is large, FAP would be low but correct detection probability (CDP) would be lower, especially when the jammer's power is low. The reason is that the jammer's RPs may not appear in a large number of estimated pilot RP sets. Finally, note that the set $\mathcal{Q}_g$ gives us an estimate of the jammer's RPs since the common RPs are jammer's paths with a high probability. 
	
	\section{Jamming Suppression}\label{JAMMING_SUP}
	In this section, we utilize the directional information attained during the jamming detection phase to design a signal detection scheme which mitigates the intentional interference of the jammer in the data transmission phase. The proposed method relies on a modified channel estimation which takes into account the effect of the jammer. This channel estimation is implemented by projecting the received pilot signals onto the orthogonal complement of the jammer’s angular subspace and  is based on a linear minimum mean squared error (LMMSE) estimation technique in the training phase. 
	Afterward, in the data transmission phase, a maximum ratio combining (MRC) is constructed by the use of the aforementioned estimated channels. We show that the combining vector is orthogonal to the jammer’s channel and can effectively suppress the effect of jamming. 
	
	\subsection{Channel Estimator}
	As discussed in section \ref{Sys_Mod}, the channel of a user is constituted of two types of information, 1) a set of active RPs, 2) the gains of these active RPs. We obtained an estimate of RPs along which pilot signals were received, $\hat{\Omega}_{k,w}$, which also included paths of the jammer. The proposed jamming detection scheme yields an estimate of the jammer's spatial signature. By subtracting the jammer's paths from $\hat{\Omega}_{k,w}$, we obtain a set that only comprises the $k$th user's RPs, i.e. $\hat{\Omega}_k=\hat{\Omega}_{k,w}\setminus\mathcal{Q}_g$. Thus, by estimating the gain along these paths, an estimation of the users' channel will be attained. 
	
	The signals received in the training phase along with $i$th RP for $i\in\hat{\Omega}_k$ can be written as
	\begin{equation}
	    [\tilde{\mathbf{y}}_{t,k}^n]_i=\mathbf{a}(\phi_i)^H\mathbf{y}_{t,k}^n=\sqrt{\tau p_{t,k}\mu_k}\tilde{g}_{k,i}^n+[\tilde{\mathbf{z}}_{t,k}^n]_i.
	\end{equation}
	Therefore, according to \cite{Kay97}, we can define the LMMSE estimation of the $i$th RP's gain as
	\begin{equation}\label{MMSE_Gains}
	    \hat{\tilde{g}}_{k,i}^{n}=\frac{\mathbb{E}\{\tilde{g}_{k,i}^n[\tilde{\mathbf{y}}_{t,k}^n]_i\}}{\mathbb{E}\{|[\tilde{\mathbf{y}}_{t,k}^n]_i|^2\}}[\tilde{\mathbf{y}}_{t,k}^n]_i=\dfrac{\sqrt{\tau p_{k,t}\mu_k}}{\sigma_z^2+\tau             p_{k,t}\mu_k}\mathbf{a}(\phi_i)^H\mathbf{y}_{t,k}^n,
	\end{equation}
	where $\mu_k=\frac{M\beta_k}{C_k}$. The mean-square of the estimated gain are
	\begin{equation}\label{MS_Gain}
	\xi_k=\mathbb{E}\{|\hat{\tilde{g}}_{k,i}^{n}|^2\}=\frac{|\mathbb{E}\{\tilde{g}_{k,i}^n[\tilde{\mathbf{y}}_{t,k}^n]_i\}|^2}{\mathbb{E}\{|[\tilde{\mathbf{y}}_{t,k}^n]_i|^2\}}=\dfrac{\tau p_{k,t}\mu_k}{\sigma_z^2+\tau p_{k,t}\mu_k}.
	\end{equation}
	Note that this estimator is the same for all the RPs of a user, since they all have the same distribution. Moreover, the estimation error is defined as $e_{k,i}^n=\hat{\tilde{g}}_{k,i}^{n}-\tilde{g}_{k,i}^{n}$ and its mean-square is $\mathbb{E}\{|e_{k,i}^n|^2\}=\dfrac{\sigma_z^2}{\sigma_z^2+\tau p_{k,t}\mu_k}$. By having the RP set of a user and the RPs' gains, the channel of the $k$th user can be estimated as
	\begin{equation}\label{Channel_Estimation}
	    \hat{\mathbf{h}}_k^n\triangleq\sqrt{\mu_k}\sum_{i\in\hat{\Omega}_k}\hat{\tilde{g}}_{k,i}^{n}\mathbf{a}(\phi_i).
	\end{equation}
	This estimation technique has a great advantage that makes the estimated channels orthogonal to the jammer's channel, i.e. $(\hat{\mathbf{h}}_k^n)^H\mathbf{h}_w^n\simeq 0$. It stems from the fact that by eliminating the jammer's active RPs, the angular subspace of the jammer's channel is excluded and the users' channels lie in its null space. Although some of the jammer's RPs which have insignificant gains may remain, especially when the jammer's power is very low, their impact on the estimation accuracy and orthogonality to the jammer's channel will be negligible due to their small gain. We should also note that for the users that have some common RPs with the jammer, the accuracy of channel estimation decreases. However, it is unlikely and would not affect the majority of users.   
	\subsection{MRC Decoder}
	As mentioned before, in this paper we adopt an MRC combining method which is a simple and efficient precoder for massive MIMO systems. Hence, in the uplink data transmission phase, the decoding matrix will be $\mathbf{V}^n=\hat{\mathbf{H}}^n$, where $\hat{\mathbf{H}}^n=\{\hat{\mathbf{h}}_1^n,\hat{\mathbf{h}}_2^n,\ldots,\hat{\mathbf{h}}_K^n\}$. As mentioned before, canceling out the jammer's RPs would eliminate its deliberate interference due to the orthogonality of the combining vectors to the jammer's channel which can be interpreted as a kind of spatial filtering.
    Using MRC, terms of SINR in (\ref{SINR}) for the $k$th user in the $n$th subcarrier will be obtained as
	\begin{align}
	    &|\mathbb{E}\{(\hat{\mathbf{h}}_k^n)^H\mathbf{h}_k^n\}|^2=\mu_k^2C_{k/w}^2\xi_k^2, \\
	    &\mathtt{var}\{(\hat{\mathbf{h}}_k^n)^H\mathbf{h}_k^n\}=\mu_k^2C_{k/w}\xi_k^2,\\
	    &\mathbb{E}\{|(\hat{\mathbf{h}}_k^n)^H\mathbf{h}_l^n|^2\}=C_{k,l}\xi_k\mu_k\mu_l,\\
	    &\mathbb{E}\{|(\hat{\mathbf{h}}_k^n)^H\mathbf{h}_w^n|^2\}\simeq 0,\\
	    &\mathbb{E}\{|(\hat{\mathbf{h}}_k^n)^H\mathbf{z}_d^n|^2\}=\mu_kC_{k/w}\xi_k\sigma_z^2,
	\end{align}
	where $C_{k/w}=card\{\hat{\Omega}_k\}$ and $C_{k,l}=C_{l,k}=card\{\hat{\Omega}_k\cap\hat{\Omega}_l\}$. The derivation of these terms is included in \textbf{Appendix \ref{APP2}} with more details. Now, after calculating these terms, the effective SINR of the $k$th user in the $n$th sub-carrier is described as
	\begin{equation}\label{SINR_MRC}
	\rho_k^{n,MRC}=\dfrac{p_{d,k}\mu_kC_{k/w}^2\xi_k}{p_{d,k}\mu_kC_{k/w}\xi_k+\sum_{\stackrel{l=1}{l\neq k}}^{K}p_{d,l}\mu_lC_{k,l}+C_{k/w}\sigma_z^2}.
	\end{equation}

	\section{Simulation Results}\label{Simulation}
	
	In this section, we evaluate the proposed jamming detection and suppression scheme by numerical simulations. In our simulations, we assumed $K=10$ and equal user transmit power in both training and data transmission phases with $p_{k,d}=p_{k,t}=0$ dBw for all users and noise power is $\sigma_z^2=-25$ dBw. Also, the large scale fading coefficients are assumed to be equal to $0$ dB for users and the jammer, i.e. $\beta_k=\beta_w=1$. The length of pilot signals is taken as $\tau=K=10$ and the length of coherence block is set to $T=200$. The users' and jammer's average incident angles are considered to be uniformly distributed in the interval $[-\pi/2+\Delta/2,\pi/2-\Delta/2]$ and angular spread is constant for all terminals.  The number of sub-carriers is $N_d=1, 20$ and the threshold in (\ref{RP_Dec_Rule}) is selected as $\epsilon_k=0.02,0.11$. 

    \begin{figure}[t!]
		\centering
		\includegraphics[width=10cm]{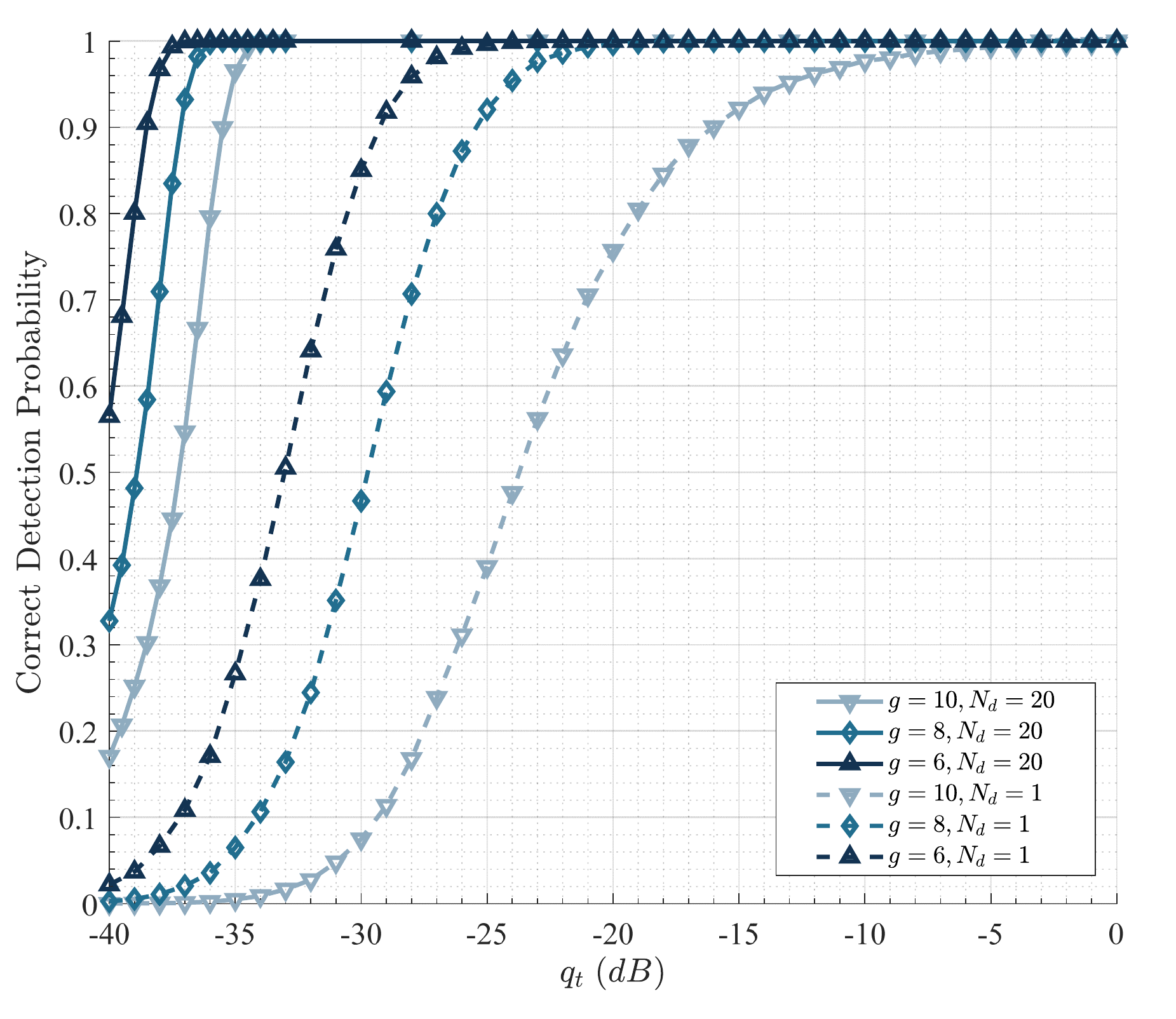}
		\caption{\small{CDP of the proposed jamming detection method versus the jammer's power in the training phase (for $M=200$ and $\Delta=\pi/18$)} } \label{CDP}
	\end{figure}
	
	At first, we evaluate the performance of the proposed jamming  detection method by evaluating its CDP and FAP. Fig. \ref{CDP} shows the CDP versus the jammer's power $q_t$ for different values of  $g=6,8,10$. It is apparent that CDP improves with the jammer's power as the jammer's RPs become easier to estimate. Furthermore, this figure shows that using a larger number of sub-carriers $N_d$ leads to a better CDP performance. For example, using $N_d=20$ sub-carriers helps the BS to detect the jammer with 10dB lower power. Besides, a lower value of $g$ results in a better detection probability. 
	
	\begin{figure}[t!]
		\centering
		\includegraphics[width=10cm]{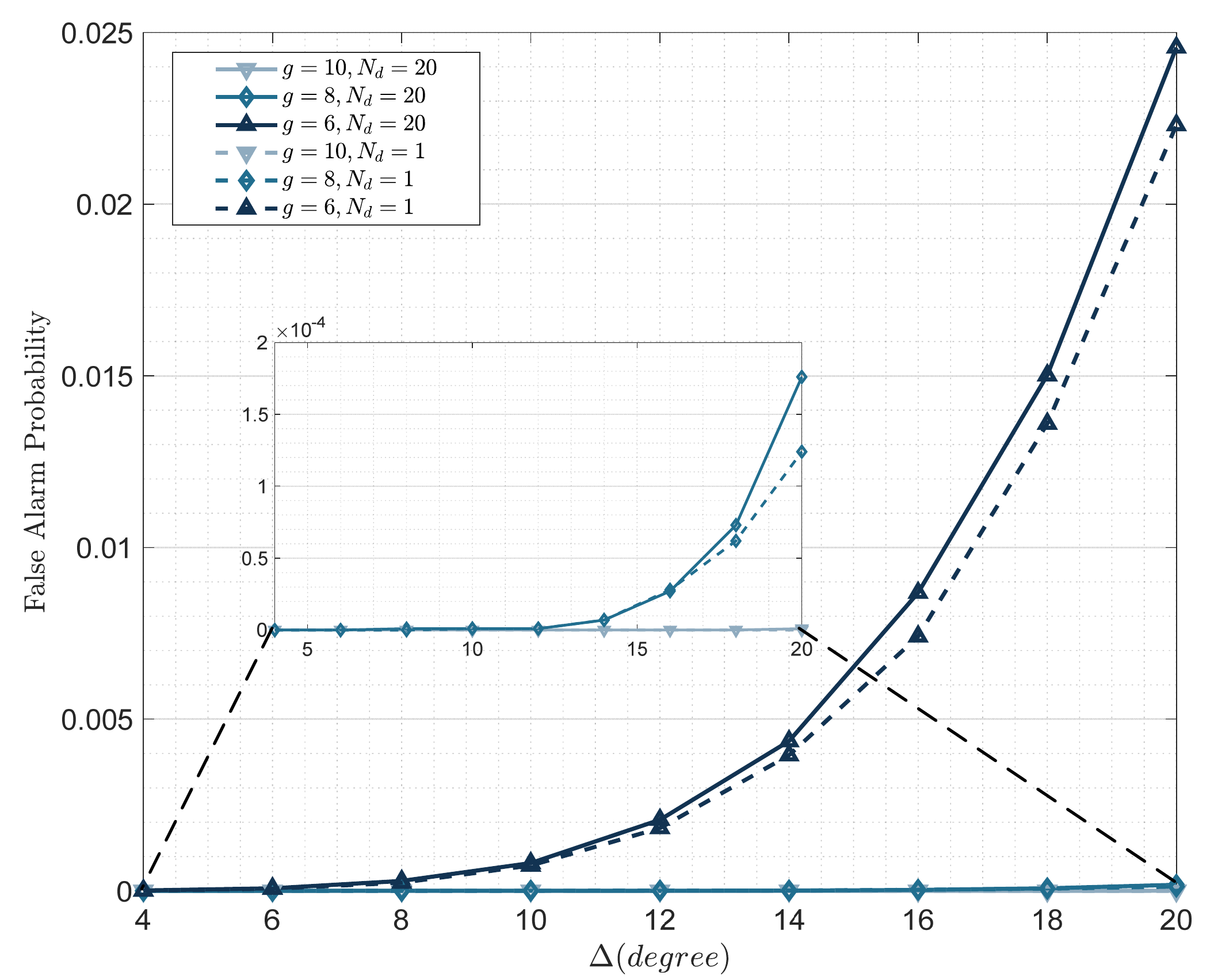}
		\caption{\small{FAP of the proposed jamming detection method versus angular spread $\Delta$  (for $M=200$)}} \label{FAP}
	\end{figure}
	
	Fig. \ref{FAP} demonstrates the relation between the FAP and angular spread $\Delta$. As shown in this figure, FAP increases with the angular spread $\Delta$. It is due to the fact that with a larger angular spread, the probability that an RP is common among a large subset of users would rise. In addition, a smaller value of $g$ would result in a higher FAP. In other words, the FAP of the proposed detection method can be decreased by a more careful selection of $g$. 
	
	\begin{figure}[t!]
		\centering
		\includegraphics[width=10cm]{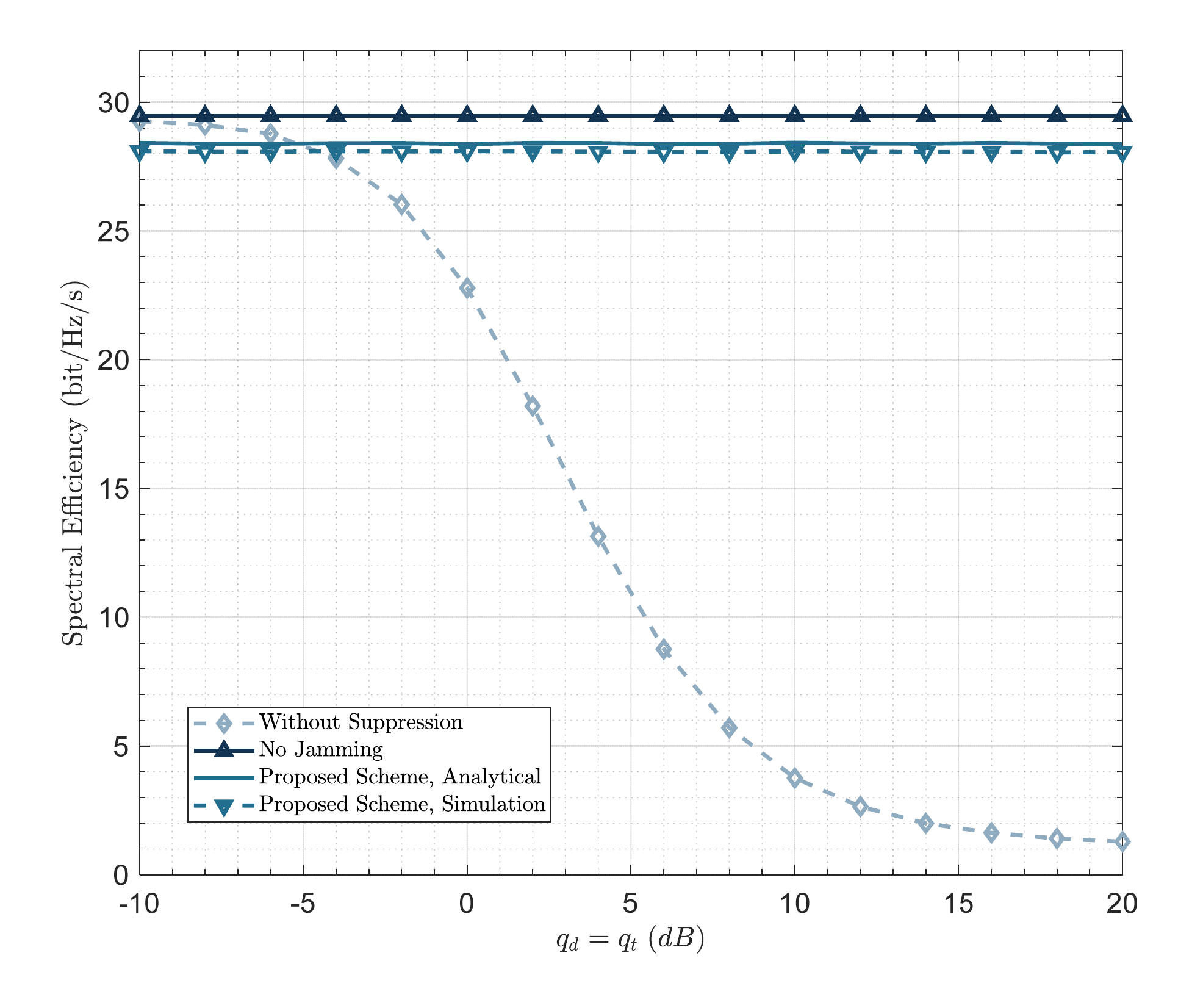}
		\caption{\small{Sum SE versus the jammer's power (for $K=10$, $M=200$, $\Delta=\pi/18$, $g=6$, $N_d=20$ and $T=200$)}} \label{SE_JP}
	\end{figure}
	
	Fig. \ref{SE_JP} depicts how the jammer's power affects sum-SE of the network for three different scenarios, i) no jammer in the network, ii) jamming with the proposed jamming detection and suppression scheme and iii) jamming when no suppression. In this simulation, we considered $M=200$, $\Delta=\pi/18$, $g=6$ and $N_d=20$. We can see that the proposed suppression technique is able to effectively cancel out the jamming attack and the resulting sum-SE  is very close to the case that there is no jammer in the network. However, it is worth mentioning that for jamming power below $-5$ dBw, using the proposed suppression scheme is undesirable. However, in this case, the effect of the jamming on the system's performance is negligible.
	
	\begin{figure}[t!]
		\centering
		\includegraphics[width=10cm]{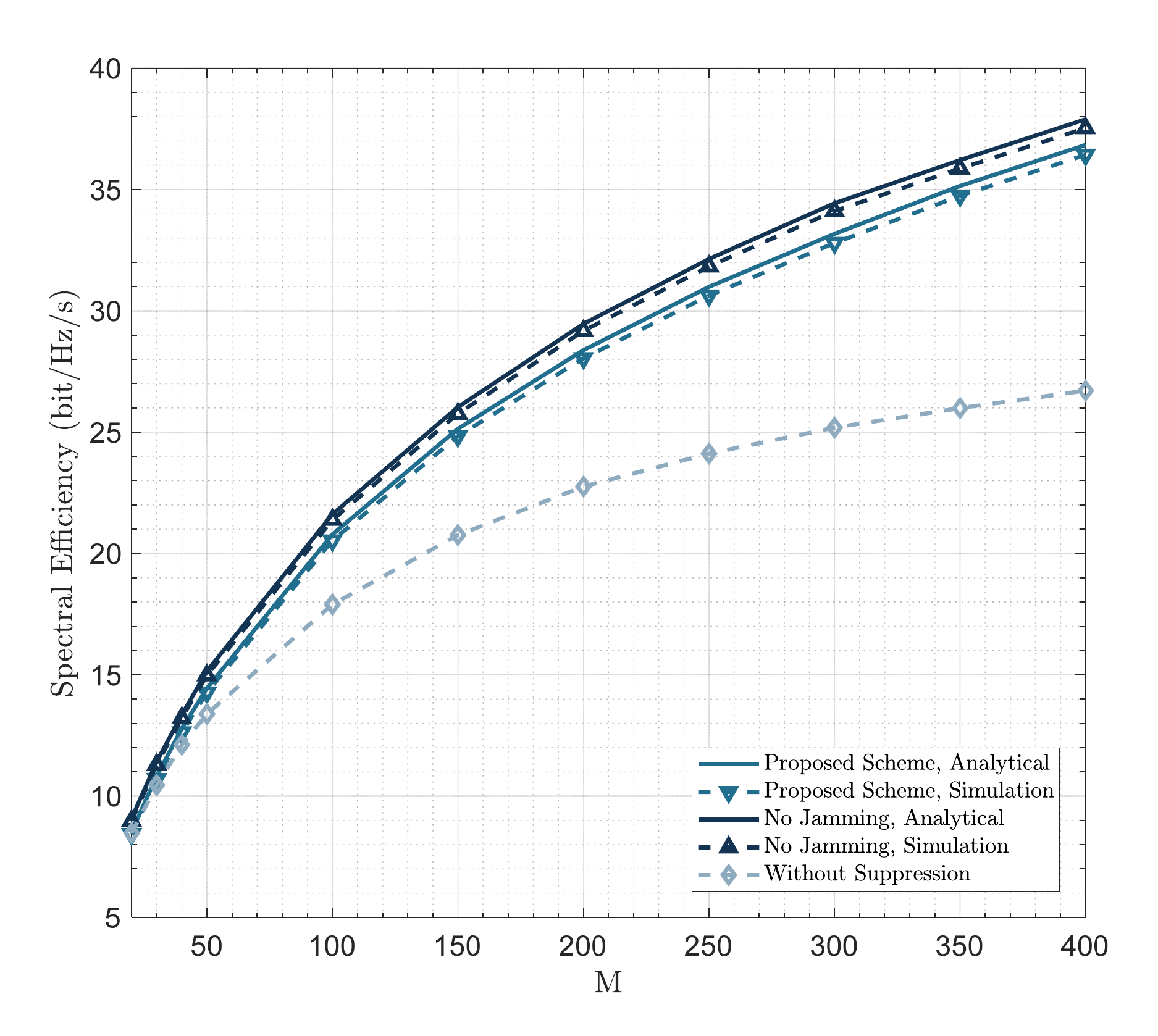}
		\caption{\small{Sum SE versus the number of BS's antenna $M$ (for $q_t=q_d=0$ dB, $K=10$,  $\Delta=\pi/18$, $g=6$, $N_d=20$ and $T=200$)}} \label{SE_Nt}
	\end{figure}

	Fig. \ref{SE_Nt} illustrates how sum-SE changes with the number of array antennas. In this figure we also consider three aforementioned scenarios and set the parameters as $q_t=q_d=0$ dB, $g=6$, $N_d=20$ and $\Delta=\pi/18$. As expected, the sum-SE of the system increases with a higher number of array antennas. It can also be deducted from this figure that the proposed strategy can successfully reject the jammer's attack and be close to the case that there is no jamming in the network. The insignificant loss of SE is the result of excluding the common RPs between the jammer and the users.
	
	\section{Conclusion}\label{Conclusion}
    In this paper, we investigated a direction (angular)-based strategy for detecting and mitigating jamming attacks on the uplink transmission of mmWave massive MIMO networks. The proposed scheme first utilized the received signals in the training phase to estimate the active RPs (i.e. directions) from which a pilot signal is received. Then, by using this information, the presence of the jammer and its direction were detected. Finally, using the information obtained in the previous step, a suppression technique was proposed which was able to countermeasure the jammer's attacks in the data transmission phase. Our numerical analysis showed the effectiveness of the proposed strategy against the jamming attacks under different scenarios.

\section{Appendix}
\subsection{\emph{Proof of} \textbf{Remark 1}}\label{APP1}
	There are $\binom{K}{g}$ combinations of $g$ user from $K$ users. We denote the set of chosen users by $\mathcal{M}_i, i=1,...,\binom{K}{g}$. Therefore, we can define
	\begin{align}\label{Eq.AP1}
		\nonumber P_g^K=&Prob(\bigcup_{i=1}^{\binom{K}{g}}(\bigcap_{j\in \mathcal{M}_i}\Omega_j\neq\emptyset))\\
		&\nonumber\stackrel{a}{\leq}\min(1,\sum_{i=1}^{\binom{K}{g}}Prob(\bigcap_{j\in \mathcal{M}_i}\Omega_j\neq\emptyset))\\
		&\stackrel{b}{=}\min (1,\binom{K}{g}Prob(\bigcap_{j\in \mathcal{M}_1}\Omega_j\neq\emptyset)),
	\end{align}
	where $\mathcal{M}_1=\{1,\ldots,g\}$ and ($a$) is derived from Frechet inequality for union of events and ($b$) holds from the fact that $Prob(\bigcap_{j\in \mathcal{M}_i}\Omega_j)$ is constant for different combinations of users since no distinctive characteristics are considered for different users. Now, we investigate the probability $Prob(\bigcap_{j=1}^{g}\Omega_j)$. First, we define the $\mathcal{I}_{int}$ as intersection of $\mathcal{I}_j,j=1,\ldots,g$. Now, we can write
	\begin{equation}\label{Eq.AP2}
		Prob(\bigcap_{j=1}^{g}\Omega_j\neq\emptyset)=P_\Psi Prob(\mathcal{I}_{int}\neq\emptyset)\stackrel{c}{\leq}Prob(\mathcal{I}_{int}\neq\emptyset),
	\end{equation}
	where $P_\Psi$ is the probability that there is at least one RP in $\mathcal{I}_{int}$ and ($c$) holds since $P_\Psi\leq 1$. In order to $\mathcal{I}_{int}$ to be non-empty, all $\mathcal{I}_k$ should overlap with each other, which would be achieved if for any $i,j\in \mathcal{M}_1, i<j$, the distance between $\bar{\theta}_i$ and $\bar{\theta}_j$ is less than $\Delta$. We define these events as
	\begin{equation}\label{Eq.AP3}
		E_{i,j}\triangleq(|\bar{\theta}_i-\bar{\theta}_j|\leq\Delta)\equiv(\mathcal{I}_i\cap\mathcal{I}_j\neq\emptyset)
	\end{equation}
	and $Prob(E_{i,j})=1-(\frac{\pi-2\Delta}{\pi-\Delta})^2$ since $\bar{\theta}_i, i=1,\ldots,K$ are independent uniformly distributed in interval $[\frac{\pi}{2}-\frac{\Delta}{2},-\frac{\pi}{2}+\frac{\Delta}{2}]$. Now, we can write
	\begin{equation}\label{Eq.AP4}
		Prob(\mathcal{I}_{int}\neq\emptyset)=Prob(\bigcap_{\stackrel{i,j=1}{i<j}}^{g}E_{i,j}).
	\end{equation}
    Although these events are not independent, but they can be divided into independent events, e.g. $E_{i,j}\indep E_{i,k},\; i,j,k=1,\ldots,g;\; j\neq k;\; i<j,k$. We define new events as intersection of independent events
	\begin{equation}\label{Eq.AP5}
		E_i=\bigcap_{j=i+1}^gE_{i,j},\;i=1,\ldots,g-1.
	\end{equation}
	These new events are not independent but their probability can be attained by $Prob(E_i)=(1-(\frac{\pi-2\Delta}{\pi-\Delta})^2)^{g-i}$. By substituting (\ref{Eq.AP5}) to (\ref{Eq.AP4}), we can obtain
	\begin{equation}\label{Eq.AP6}
		Prob(\bigcap_{\stackrel{i,j=1}{i<j}}^{g}E_{i,j})=Prob(\bigcap_{i=1}^{g-1}E_i)\stackrel{d}{\leq}\min_i(Prob(E_i)),
	\end{equation}
	where ($d$) is derived according to Frechet inequality for intersection of events and $\min_i(Prob(E_i))=Prob(E_1)=(1-(\frac{\pi-2\Delta}{\pi-\Delta})^2)^{g-1}$ since it has the largest exponent. Thus, according to (\ref{Eq.AP1}-\ref{Eq.AP6}), the upper bound of (\ref{Eq12}) is derived.

	\subsection{The derivation of SINR terms using MRC decoder}\label{APP2}
	When MRC is utilized, the terms of achievable rate mentioned in (\ref{SINR}) can be evaluated as follows.
	\begin{itemize}
		\item The desired signal $|\mathbb{E}\{(\hat{\mathbf{h}}_k^n)^H\mathbf{h}_k^n|\Psi\}|^2$: 
		By substituting \eqref{Channel_Model_User_sum} and \eqref{Channel_Estimation} in $\mathbb{E}\{(\hat{\mathbf{h}}_k^n)^H\mathbf{h}_k^n|\Psi\}$, we have
		\begin{align}
		    \nonumber\mathbb{E}\{(\hat{\mathbf{h}}_k^n)^H\mathbf{h}_k^n|\Psi\}&=\mathbb{E}\{(\sqrt{\mu_k}\sum_{i\in\hat{\Omega}_k}\hat{\tilde{g}}_{k,i}^{n}\mathbf{a}(\phi_i))^H\sqrt{\mu_k}\sum_{i\in\Omega_k}\tilde{g}_{k,i}^{n}\mathbf{a}(\phi_i)\}\\\label{App2_Eq1}&\stackrel{e}{=}\mu_k\mathbb{E}\{ \sum_{i\in\hat{\Omega}_k} {\hat{\tilde{g}}_{k,i}^{n}}^* \tilde{g}_{k,i}^{n} \},
		\end{align}
		where ($e$) holds since $a(\phi_i)^Ha(\phi_j)=0$ for $i\neq j$ and $\hat{\Omega}_k\subseteq \Omega_k$. We know  $e_{k,i}^n=\hat{\tilde{g}}_{k,i}^{n}-\tilde{g}_{k,i}^{n}$, and hence, we can rewrite the \eqref{App2_Eq1} as
		\begin{align}
            \nonumber\mathbb{E}\{(\hat{\mathbf{h}}_k^n)^H\mathbf{h}_k^n|\Psi\}&=\mu_k\mathbb{E}\{ \sum_{i\in\hat{\Omega}_k} |\hat{\tilde{g}}_{k,i}^{n}|^2 \}- \mu_k\mathbb{E}\{\sum_{i\in\hat{\Omega}_k} {\hat{\tilde{g}}_{k,i}^{n}}^* e_{k,i}^n\}\\ \label{App2_Eq2} &\stackrel{f}{=} \mu_k\mathbb{E}\{ \sum_{i\in\hat{\Omega}_k} |\hat{\tilde{g}}_{k,i}^{n}|^2 \},
        \end{align}
        where ($f$) holds since estimated RP gain $\hat{\tilde{g}}_{k,i}^{n}$ is uncorrelated with the estimation error $e_{k,i}^n$ in LMMSE \cite{Kay97}, i.e. $\mathbb{E}\{ {\hat{\tilde{g}}_{k,i}^{n}}^* e_{k,i}^n \}=0$. The estimated gain $\hat{\tilde{g}}_{k,i}^{n}$ is a Gaussian variable with a known variance $\xi_k$, thus $\sum_{i\in\hat{\Omega}_k} |\hat{\tilde{g}}_{k,i}^{n}|^2\thicksim Gamma(C_{k/w},\xi_k)$. Therefore
        \begin{equation}\label{App2_Eq3}
             \mathbb{E}\{ \sum_{i\in\hat{\Omega}_k} |\hat{\tilde{g}}_{k,i}^{n}|^2\}=C_{k/w}\xi_k,
        \end{equation}
        Considering \eqref{App2_Eq2} and \eqref{App2_Eq3}, we can conclude $\mathbb{E}\{(\hat{\mathbf{h}}_k^n)^H\mathbf{h}_k^n|\Psi\}=\mu_kC_{k/w}\xi_k$. Therefore, 
        \begin{equation}\label{App2_Eq4}
            |\mathbb{E}\{(\hat{\mathbf{h}}_k^n)^H\mathbf{h}_k^n|\Psi\}|^2=\mu_k^2C_{k/w}^2\xi_k^2.
        \end{equation}

		\item The signal gain uncertainty $\mathtt{var}\{(\hat{\mathbf{h}}_k^n)^H\mathbf{h}_k^n\}$:
		As mentioned above $\sum_{i\in\hat{\Omega}_k} |\hat{\tilde{g}}_{k,i}^{n}|^2\thicksim Gamma(C_{k/w},\xi_k)$, hence
		\begin{equation}\label{App2_Eq5}
			\mathtt{var}\{(\hat{\mathbf{h}}_k^n)^H\mathbf{h}_k^n\}=\mu_k^2\mathtt{var}\{\sum_{i\in\hat{\Omega}_k}|\hat{\tilde{g}}_{k,i}^{n}|^2\}=\mu_k^2C_{k/w}\xi_k^2.
		\end{equation}
		\item Interference exerted by other users $\mathbb{E}\{|(\hat{\mathbf{h}}_k^n)^H\mathbf{h}_l^n|^2\}$:
		\begin{equation}\label{App2_Eq6}
			\mathbb{E}\{|(\hat{\mathbf{h}}_k^n)^H\mathbf{h}_l^n|^2\}=\mathbb{E}\{((\mathbf{h}_l^n)^H\hat{\mathbf{h}}_k^n)((\hat{\mathbf{h}}_k^n)^H\mathbf{h}_l^n)\}
		\end{equation}
		By substituting $\hat{\mathbf{h}}_k^n=\sqrt{\mu_k}\sum_{i\in\hat{\Omega}_k}\hat{\tilde{g}}_{k,i}^{n}\mathbf{a}(\phi_i)$ and $\mathbf{h}_l^n=\\\sqrt{\mu_l}\sum_{i\in\Omega_l}\tilde{g}_{l,i}^{n}\mathbf{a}(\phi_i)$ in (\ref{App2_Eq6}), we obtain
		\begin{align}\label{App2_Eq7}
			&\nonumber\mu_k\mu_l\mathbb{E}\{(\sum_{i\in\Omega_{k,l}}\tilde{g}_{l,i}^{n^*}\hat{\tilde{g}}_{k,i}^n)(\sum_{i\in\Omega_{k,l}}\hat{\tilde{g}}_{k,i}^{n^*}\tilde{g}_{l,i}^n)\}
			\\&\stackrel{g}{=}\mu_k\mu_l\sum_{i\in\Omega_{k,l}}\mathbb{E}\{|\hat{\tilde{g}}_{k,i}^n|^2|\tilde{g}_{l,i}^n|^2\},
		\end{align}
		where the equality ($g$) holds since $\mathbb{E}\{\hat{\tilde{g}}_{k,i}^n\hat{\tilde{g}}_{k,j}^{n^*}\}=0,\; i\neq j$, $\mathbb{E}\{\tilde{g}_{l,i}^n\tilde{g}_{l,j}^{n^*}\}=0,\; i\neq j$ and the fact that $\hat{\tilde{g}}_{k,i}^n$ is uncorrelated with $\tilde{g}_{l,i}^n$. Therefore,
		\begin{align}\label{App2_Eq8}
			\nonumber\mathbb{E}\{|(\hat{\mathbf{h}}_k^n)^H\mathbf{h}_l^n|^2\}&=\mu_k\mu_l\sum_{i\in\Omega_{k,l}}\mathbb{E}\{|\hat{\tilde{g}}_{k,i}^n|^2|\}\mathbb{E}\{|\tilde{g}_{l,i}^n|^2\}\\
			&=C_{k,l}\xi_k\mu_k\mu_l.
		\end{align}
		\item Interference exerted by the jammer $\mathbb{E}\{|(\hat{\mathbf{h}}_k^n)^H\mathbf{h}_w^n|^2\}$: 
		As mentioned in section \ref{JAMMING_SUP}, since the subspace of the jammer is excluded from estimated channels, they would be approximately orthogonal. Thus
		\begin{equation}\label{App2_Eq9}
			\mathbb{E}\{|(\hat{\mathbf{h}}_k^n)^H\mathbf{h}_w^n|^2\}\simeq 0.
		\end{equation}
		
		\item The noise term $\mathbb{E}\{|(\hat{\mathbf{h}}_k^n)^H\mathbf{z}_d^n|^2\}$:
		\begin{equation}\label{App2_Eq10}
			\mathbb{E}\{|(\hat{\mathbf{h}}_k^n)^H\mathbf{z}_d^n|^2\}\stackrel{h}{=}\mathbb{E}\{||\hat{\mathbf{h}}_k^n||^2\}\mathbb{E}\{||\mathbf{z}_d^n||^2\}=\mu_kC_{k/w}\xi_k\sigma_z^2,
		\end{equation}
		where the first equality ($h$) is correct since noise and the estimated channel are uncorrelated. 
	\end{itemize}Finally, by substituting (\ref{App2_Eq4}), (\ref{App2_Eq5}), (\ref{App2_Eq8}), (\ref{App2_Eq9}) and (\ref{App2_Eq10}) in (\ref{SINR}), we can obtain (\ref{SINR_MRC}).

\bibliographystyle{ieeetr}
\bibliography{main}
\end{document}